\documentclass[aps,showpacs,preprint,tightenlines,nofootinbib,superscriptaddress]{revtex4}
\usepackage{amsmath2000,amssymb,graphicx}

\newcommand{\ket}[1]{\lvert#1\rangle}
\newcommand{\bra}[1]{\langle#1\rvert}
\newcommand{\sB}{\sigma_\mathrm{b}}
\newcommand{\sS}{\sigma_\mathrm{s}}
\newcommand{\m}{\hphantom{-}}
\newcommand{\phd}{{\vphantom{\dagger}}}
\newcommand{\ds}{\displaystyle}
\newcommand{\ts}{\textstyle}
\newcommand{\half}{\ts\frac{1}{2}}
\newcommand{\om}{\omega}
\newcommand{\pt}{\partial}
\renewcommand{\L}{\ensuremath{\mathcal{L}}}
\newcommand{\beq}{\begin{equation}}
\newcommand{\eeq}{\end{equation}}
\newcommand{\eeql}[1]{\label{#1}\end{equation}}
\newcommand{\bea}{\begin{eqnarray}}
\newcommand{\eea}{\end{eqnarray}}
\newcommand{\eeal}[1]{\label{#1}\end{eqnarray}}
\newcommand{\hn}{\mskip-0.5\thinmuskip}
\newcommand{\hp}{\mskip0.5\thinmuskip}
\DeclareMathOperator{\Tr}{Tr}
\DeclareMathOperator{\re}{Re}
\DeclareMathOperator{\im}{Im}

\begin{document}

\title{Liouville invariance in quantum and classical mechanics}

\author{Alec \surname{Maassen van den Brink}}
\email{alec@dwavesys.com}
\affiliation{D-Wave Systems Inc., 320-1985 West Broadway, Vancouver, BC, V6J 4Y3, Canada}
\author{A.M. Zagoskin}
\affiliation{D-Wave Systems Inc., 320-1985 West Broadway, Vancouver, BC, V6J 4Y3, Canada}
\affiliation{Physics and Astronomy Dept., The University of British Columbia, 6224 Agricultural Rd., Vancouver, BC, V6T 1Z1, Canada}

\date{\today}

\begin{abstract}
The density-matrix and Heisenberg formulations of quantum mechanics follow---for unitary evolution---directly from the Schr\"odinger equation. Nevertheless, the symmetries of the corresponding evolution operator, the Liouvillian $\L=i[\,\boldsymbol{\cdot}\,,H]$, need not be limited to those of the Hamiltonian~$H$. This is due to \L\ only involving eigenenergy \emph{differences}, which can be degenerate even if the energies themselves are not. Remarkably, this possibility has rarely been mentioned in the literature, and never pursued more generally. We consider an example involving mesoscopic Josephson devices, but the analysis only assumes familiarity with basic quantum mechanics. Subsequently, such \emph{\L-symmetries} are shown to occur more widely, in particular also in classical mechanics. The symmetry's relevance to dissipative systems and quantum-information processing is briefly discussed.
\end{abstract}


\maketitle

\section{Introduction}

Density matrices streamline quantum statistics, by combining the calculation of quantum-mechanical expectation values and classical averaging over a probability distribution of states, into a single trace operation~\cite{messiah,blum}. The aesthetically inclined might further appreciate that for a pure state $\ket{\psi}$, its density matrix $\ket{\psi}\bra{\psi}$ is uniquely defined, unlike $\ket{\psi}$ itself~\cite{weyl}.

While the state vector obeys the Schr\"odinger equation $id_t\ket{\psi}=H\ket{\psi}$ ($\hbar=1$), the density matrix evolves according to the von Neumann (or quantum Liouville) equation
\beq
  d_t\rho=i[\rho,H]\equiv\L(\rho)\;,
\eeql{neumann}
where \L\ is the Liouville (super-)operator. Because \L\ is defined in terms of the Hamiltonian~$H$, the two operators are closely related; in particular, any symmetry of $H$ is inherited by \L\ (see Section~\ref{theory} for a precise statement). The density-matrix formulation is also uniquely useful in more general cases where the Liouvillian no longer is given by $\L=i[\,\boldsymbol{\cdot}\,,H]$~\cite{lindblad,louisell}, or where the evolution even is effectively nonlocal in time~\cite{weiss}. In such situations, a pure state $\rho=\rho^2$ will in general evolve into a mixture, so that a Schr\"odinger formulation is not possible. This occurs when the system being studied is coupled to a ``bath,'' of which complete knowledge is neither possible nor desirable. Tracing out the bath degrees of freedom then yields the \emph{reduced density matrix} of the system. The latter's effective evolution can deviate from unitarity by (a) dissipation (with the attendant fluctuations~\cite{KTH}): changes in the diagonal (in the energy basis) elements of $\rho$ representing occupation probabilities, and (b)~dephasing: random fluctuations of the system's energy levels, which suppress off-diagonal elements of $\rho$ even without bath-induced dissipative transitions between those levels. Studying and controlling the non-unitary evolution of $\rho$ is for instance crucial if its  off-diagonal elements are used to store (quantum) information~\cite{N&C}.

For the moment, however, we restrict ourselves to unitary evolution (\ref{neumann}). Then, the Schr\"odinger and von Neumann equations would seem to be equivalent, a complete set of solutions for the latter being given by
\beq
  \ket{\psi_j}\bra{\psi_k}\exp\{i(\om_k{-}\om_j)t\}\;,
\eeql{L-ev}
where $H\ket{\psi_j}=\om_j\ket{\psi_j}$.\footnote{We will not dwell on the mathematical subtleties which arise in the case of continuous spectra.} It is our purpose to highlight the fact that the reverse need \emph{not} hold, and to examine the consequences. Namely, consider a system such that
\beq
  \om_3-\om_1=\om_4-\om_2\neq0\;.
\eeql{om-diff}
For any $a,b$, the superposition $[a\ket{\psi_1}\bra{\psi_2}+b\ket{\psi_3}\bra{\psi_4}]\exp\{i(\om_2{-}\om_1)t\}$ is then a solution of~(\ref{neumann}), which is not of the form (\ref{L-ev}) if $ab\neq0$. In other words, the concept of degeneracy is wider for the von Neumann than for the Schr\"odinger equation, since only energy \emph{differences} have to be equal \cite{B&P}---physically appealing, as it is only those differences which are gauge invariant and, hence, observable. The generalization to more than four states will be obvious.

\section{Formalism of Liouville symmetry}
\label{theory}

Before discussing specific instances, let us outline the theory of the above-mentioned \emph{Liouville symmetry} or \emph{\L\nobreakdash-symmetry}, in a form general enough to apply to classical mechanics as well. First of all, a few notions from operator theory should be recapitulated, and some notation established. However, readers familiar with superoperators, or on the contrary, interested more in applications than in formalism, can skip to Section~\ref{circuit} and consult the present section as needed.

Quantum mechanics deals with operators on a Hilbert space of states $\mathbb{H}$. These operators can themselves be seen as points in a \emph{Liouville space} $\mathbb{L}$~\cite{EBW}, which can be given the inner product
\beq
  \langle A,B\rangle\equiv\Tr\{A^{\dagger}B\}\;.
\eeql{superprod}
As always, the trace can be calculated in any basis, though it has the form $\sum_j\bra{\psi_j}\boldsymbol{\cdot}\ket{\psi_j}$ only in orthonormal ones. Sidestepping the issue that the trace may well diverge in infinite dimensions, the inner-product axioms are readily verified. In particular, there are no problems with density matrices, which are positive-semidefinite operators normalized according to $\langle\rho,\openone\rangle=1$. \emph{Observables} are Hermitian operators $A$, so that $\langle A\rangle\equiv\langle\rho,A\rangle\in\mathbb{R}$.

Suppose the subspace $\mathcal{V}\subseteq\mathbb{L}$ is \emph{$\dagger$-closed}: $\mathcal{V}=\mathcal{V}^{\dagger}$. Then $\mathcal{W}\equiv\{\rho\in\mathcal{V}:\rho=\rho^{\dagger}\}$ is a real subspace spanning~$\mathcal{V}$; in particular, observables have real products (\ref{superprod}). An orthonormal basis of $\mathcal{W}$ then is a Hermitian ($\mathbb{C}$-)basis for $\mathcal{V}$; cf.~(\ref{Lmat}) below. To proceed, consider $\mathcal{V}\mathbb{H}\subseteq\mathbb{H}$. Counterintuitively, this need not itself be a subspace; cf.\footnote{Foregoing Hermiticity, there are also 3D counterexamples.}
\beq
  \begin{pmatrix} 1 & \m0 & 0 & 0\\ 0 & -1 & 0 & 0\\ 0 & \m0 & 0 & 0\\ 0 & \m0 & 0 & 0 \end{pmatrix}
  \begin{pmatrix} 1 \\ 0 \\ 0 \\ 0 \end{pmatrix} +
  \begin{pmatrix} 0 & 0 & 0 & 1 \\ 0 & 0 & 0 & 1 \\ 0 & 0 & 1 & 0 \\ 1 & 1 & 0 & 0 \end{pmatrix}
  \begin{pmatrix} 0 \\ 0 \\ 1 \\ 0 \end{pmatrix}\;,
\eeq
which is not in the range of any matrix in the span of the two indicated ones. If however $\mathcal{V}\mathbb{H}$ in fact is a subspace of $\mathbb{H}$, one may define its projector $P\equiv P_{\mathcal{V}\mathbb{H}}$. If furthermore one has $\mathcal{V}=P\mathbb{L}P$, we call $\mathcal{V}$ a \emph{conventional} Liouville subspace, generated by the Hilbert subspace~$\mathcal{V}\mathbb{H}$. In the notation developed below, $\mathcal{P}=P_\mathrm{l}P_\mathrm{r}$ for the projector onto $\mathcal{V}$. In all other cases, $\mathcal{V}$ will be said to be an \emph{unconventional} Liouville subspace. Note that the dimension of a conventional Liouville subspace is a square, and that a proper Liouville subspace containing an invertible element is always unconventional; compare both with~(\ref{Lmat}) in Section~\ref{liouville}.

The logical next step is to consider linear \emph{superoperators}\footnote{This term here merely means ``operators acting on operators,'' and has nothing to do with conversion between bosons and fermions. Well, almost nothing---see p.~\pageref{super}.} $\mathcal{S}$ in Liouville space~\cite{crawford}. Ordinary operators $A$ can be promoted to superoperators in several ways. Presently we only introduce $A_\mathrm{l}(\rho)\equiv A\rho$ and $A_\mathrm{r}(\rho)\equiv\rho A$~\cite{B&P,EBW}, so that (\ref{neumann}) amounts to
\beq
  \L=i(H_\mathrm{r}-H_\mathrm{l})\;.
\eeql{HlHr}
\emph{A superoperator $\mathcal{S}$ is a left (right) multiplication iff $\forall A,B:\mathcal{S}(A)B=\mathcal{S}(AB)$ [$B\mathcal{S}(A)=\mathcal{S}(BA)$]. In this case,\nocorr} $\mathcal{S}=\mathcal{S}(\openone)_\mathrm{l}$ \emph{[\nocorr}$\mathcal{S}=\mathcal{S}(\openone)_\mathrm{r}$\emph{].} For the proof of the reverse implication, note that $\mathcal{S}(\openone)_\mathrm{l}(A)=\mathcal{S}(\openone)A=\mathcal{S}(\openone A)$. The other parts are analogous and/or trivial. Elementary calculation rules are $[A_\mathrm{l},B_\mathrm{l}]={[A,B]}_\mathrm{l}$ and $[A_\mathrm{r},B_\mathrm{r}]={[B,A]}_\mathrm{r}$, while $[A_\mathrm{l},B_\mathrm{r}]=0$; also, $(A_\mathrm{l/r})^{\dagger}={(A^{\dagger})}_\mathrm{l/r}$ so that one can write $A^{\dagger}_\mathrm{l/r}$ without confusion, and $\openone_\mathrm{l}=\openone_\mathrm{r}=\openone$.

In this language, the quantum-mechanical superposition principle and probability conservation imply that time evolution should have the form $\rho(t)=\mathcal{U}(t,t')[\rho(t')]$ for $t>t'$. This leads to the theory of completely positive trace-preserving superoperators~\cite{A&L,barbara}.
\begin{quote}
In its full scope, this paper advocates studying the symmetries of $\mathcal{U}$ in Liouville space, in any physics problem.
\end{quote}
Explicitly, symmetries are superoperators $\mathcal{S}$ commuting with $\mathcal{U}$,
\beq
  [\mathcal{U},\mathcal{S}]=0\;.
\eeql{U-S}
Of course, the commutant of $\mathcal{U}$ can always be written down formally in terms of the latter's Jordan normal form~\cite{A&L}. The point is to first find symmetries on physical grounds, and then to use these in the spectral analysis of $\mathcal{U}$ instead of relying on brute force. This conceptually parallels the case of conventional Hamiltonian symmetries, though technical aspects may well differ.

Specializing to systems without memory (at most Ohmic damping) leads to a differential formulation\footnote{In e.g.\ quantum optics, one often speaks of a \emph{master equation}.} $\dot{\rho}=\L(t)\hp(\rho)$, and the symmetries can be studied on the level of the generator\footnote{Also at this level, there is a correspondence between Schr\"odinger [${\langle A\rangle}^{\!\cdot}=\langle\L(\rho),A\rangle$] and Heisenberg [${\langle A\rangle}^{\!\cdot}=\langle\rho,\L^{\dagger}(A)\rangle$] pictures.}~\L, hence the name Liouville symmetries. If further the system is closed then \L\ is given by (\ref{HlHr}), regardless of whether \L\ and $H$ are time dependent or not. The evolution in Hilbert space is now unitary, and \L\ is readily checked to be anti-Hermitian,
\beq
  \langle A,\L(B)\rangle=-\langle\L(A),B\rangle\;.
\eeql{anti-h}

When time evolution is generated by a Hamiltonian, the question arises how the symmetries of \L\ relate to those of $H$: operators\footnote{Note that $A$ need not be Hermitian, cf.\ angular-momentum raising/lowering operators for spherical systems, and most unitary operators.} $A$ such that $[H,A]=0$. \emph{One has $[\L,\mathcal{S}]=0$ for a superoperator $\mathcal{S}$ of the form $\mathcal{S}=A_\mathrm{l}$ or $\mathcal{S}=A_\mathrm{r}$, and an \L\ of the form~(\ref{HlHr}), iff $[H,A]=0$.} In other words, an ordinary operator can be an \L-symmetry only if it is an ordinary symmetry in the first place. In fact, by~(\ref{HlHr}) and the calculation rules, $[\L,A_\mathrm{l/r}]=i{[A,H]}_\mathrm{l/r}$ is immediate. One can try to extend this result to more general symmetries $\mathcal{S}$. However, while $[H_\mathrm{l},\mathcal{S}]=[H_\mathrm{r},\mathcal{S}]=0\Rightarrow[\L,\mathcal{S}]=0$ by (\ref{HlHr}), the reverse does not hold (\textit{v.i.}). Clearly, \L\nobreakdash-symmetry is \emph{not} conveniently studied in terms of the Hamiltonian, even for closed systems---commutation with both $H_\mathrm{l}$ and $H_\mathrm{r}$ is too strong, while commutation with only one is too weak because e.g.\ $[H_\mathrm{l},A_\mathrm{r}]=0$ does not imply any symmetry.

To give some credit to $H$, its very existence will in general lead to simple \L-symmetries. Namely, in terms of the eigenstates $H\ket{\psi_j}=\om_j\ket{\psi_j}$, we can first of all define the projector $\mathcal{P}_{jk}(\rho)\equiv P_j\rho P_k$ (i.e., $\mathcal{P}_{jk}=P_{j,\mathrm{l}}P_{k,\mathrm{r}}$), where $P_{i}=\ket{\psi_i}\bra{\psi_i}$ are ordinary Hilbert-space eigenprojections.\footnote{Projectors corresponding to degenerate eigenspaces of $H$ are readily written down as a sum of such $\mathcal{P}_{jk}$. For instance, $\mathcal{P}_{\!\!_H}\equiv\sum_{\om_j=\om_k}\mathcal{P}_{jk}$ performs an energy measurement.} Note that $\mathcal{P}_{jk}$ is an \L-symmetry and commutes with both $H_\mathrm{l}$ and $H_\mathrm{r}$ (immediate from the calculation rules), but is not an ordinary operator. Slightly more interestingly, one has the \emph{transfer} $\mathcal{T}_{jk}(\rho)=\ket{\psi_k}\bra{\psi_j}\rho\ket{\psi_j}\bra{\psi_k}$, satisfying $[\L,\mathcal{T}_{jk}]=0$---any mixture of eigenstates is stationary. On the other hand, $[H_\mathrm{l/r},\mathcal{T}_{jk}]=0$ iff $\om_j=\om_k$. By definition, it further follows that \L-symmetry carries one ordinary symmetry to another. That is, if $[H,A]=0$ then also $[H,\mathcal{S}(A)]=0$. Of course, there is no guarantee that anything nontrivial ensues for any given system, i.e., $\mathcal{S}(A)$ could well be in the span of $\{\openone,H,A\}$.

We call a superoperator $\mathcal{S}$ \emph{real} if it maps Hermitian $\rho$ to Hermitian $\rho'$. From this it follows that $\mathcal{S}(\rho^{\dagger})=[\mathcal{S}(\rho)]^{\dagger}$ for general $\rho$. The evolution operators \L\ or $\mathcal{U}$ are necessarily real for any physical system (in fact, stronger conditions hold); on the other hand, $i\L$ is only Hermitian for unitary evolution. In contrast, $A_\mathrm{l}$ is Hermitian for any observable $A$ by the calculation rules, but real only if $A=a\openone$, $a\in\mathbb{R}$; apparently, the two are independent concepts. Note now that, e.g., the theorem relating \L- and $H$\nobreakdash-symmetries would not have come out so neat if one had considered $\mathcal{S}=A_\mathrm{l}^\phd A_\mathrm{r\vphantom{l}}^{\dagger}$ instead of $\mathcal{S}=A_\mathrm{l/r}$, cf.\ the examples of Section~\ref{further}. Thus, we have not demanded our \L-symmetries to be real. This seems reasonable, given that eigenvectors (\ref{L-ev}) of \L\ are non-Hermitian if they correspond to a nonzero eigenvalue, cf.\ the complex-conjugate eigenmodes of the underdamped classical harmonic oscillator. That is, one cannot formulate the whole theory in terms of Hermitian $\rho$ only.

The above abstraction from ``degenerate energy differences'' to ``commutation with \L'' is good for, if nothing else, arriving at the classical limit. Classical mechanics is defined on an even-dimensional phase-space manifold $\Gamma$~\cite{Cmech}, on which observables $f$ are real functions. Technically, it is optimal to start from a symplectic 2-form. For our purpose, however, it is sufficient to have a \emph{Poisson bracket}\footnote{There will be no confusion between the Poisson bracket and the anticommutator, since the former (latter) only arises in classical (quantum) mechanics.} $f,g\mapsto\{f,g\}$, and local \emph{canonical coordinates} $(p,q)\equiv(p_1,\ldots,p_n,q_1,\ldots,q_n)$ in terms of which
\beq
  \{f,g\}=\frac{\pt f}{\pt q_j}\frac{\pt g}{\pt p_j}-\frac{\pt f}{\pt p_j}\frac{\pt g}{\pt q_j}\;.
\eeql{poisson}
The Poisson bracket is then verified to satisfy the Jacobi identity, and one can further define the phase-space volume $dp\hp dq$, which is invariant\footnote{Even the orientation of phase-space volume is preserved.} under any canonical transformation, i.e., $(p,q)\mapsto(p',q')$ preserving the form of~(\ref{poisson}). Define the inner product $\langle f,g\rangle\equiv\int_\Gamma f^*g\,dp\hp dq$, and a density $\rho$ as a positive phase-space function. Then, normalization of $\rho$ and the calculation of expectation values proceed exactly as in the quantum case.

By analogy with the preceding, superoperators are defined as operators in the function space of observables. Given an observable $f$, the obvious associated superoperator reads $g\mapsto fg$. Because of the commutativity of classical mechanics, there is no need to notationally distinguish the two roles of $f$. \emph{A superoperator $\mathcal{S}$ corresponds to an observable iff $\mathcal{S}(f)g=\mathcal{S}(fg)$, in which case\nocorr} $\mathcal{S}=\mathcal{S}\openone$\emph{.} The proof is obvious by comparing to the quantum case.

Another important class of superoperators is derived from mappings $A:\Gamma\rightarrow\Gamma$, which we assume to be invertible. Then we can define\footnote{Note a difference with the quantum case, where the ``position'' part of such a mapping would induce an \emph{ordinary} operator $\psi(q)\mapsto\psi\boldsymbol{(}A^{-1}(q)\boldsymbol{)}$. For instance, parity is a quantum but not a classical observable.} $g\mapsto\mathcal{A}g\equiv g\circ A^{-1}$. Such a map is said to be canonical if the associated coordinate map is a canonical transformation. More elegant is the equivalent invariant criterion $\forall f,g:\{\mathcal{A}f,\mathcal{A}g\}=\mathcal{A}\{f,g\}$. \emph{A superoperator $\mathcal{S}$ corresponds to a phase-space map iff the homomorphic property $\mathcal{S}(f)\hp\mathcal{S}(g)=\mathcal{S}(fg)$ holds, and moreover $\mathcal{S}$ is invertible.} Note that the map being canonical is a separate condition. To prove this, one can reconstruct the mapping by taking sequences of functions approaching a $\delta$-function on phase space, and study their behavior under $\mathcal{S}$; we will not dwell on the technical details.\footnote{Indeed, the whole phase space can be reconstructed from the observable algebra~\cite{rudin}, so that a superoperator preserving this algebra has to preserve phase space.}

Hamiltonian dynamics is generated by an observable $H$, according to $\pt_t\rho=\{\hn H,\rho\}\equiv\L\rho$. An observable $f$ is said to be an \emph{integral} of $H$ if $\{\hn H,f\}=0$. A canonical map $A$ is said to be a \emph{Hamiltonian symmetry} if $\mathcal{A}H=H$.\footnote{Note that the invariance $\mathcal{A}(H)=H$ did not play a role in the quantum case}$^,$\footnote{It is well known that a smooth family of such symmetries generates an integral of $H$---in effect a finite-dimensional Noether charge. However, there does not seem to be a counterpart for \L-symmetries, since these are \emph{linear} superoperators already so that taking the infinitesimal generator of a smooth family gives nothing new. Compare to \emph{quantum} Hamiltonian symmetries $[H,A(\lambda)]=0$, where passing to the generator, $[H,\partial_\lambda A|_{\lambda=0}]=0$, merely yields another Hamiltonian symmetry.} As in the quantum case, a superoperator $\mathcal{S}$ is said to be an \L-symmetry if $[\L,\mathcal{S}]=0$. \emph{An observable is an \L-symmetry iff it is an integral of $H$; a canonical map generates an \L-symmetry iff it is a Hamiltonian symmetry.} For a proof, it suffices to write out $[\L,f]=\{\hn H,f\}$ and $[\L,\mathcal{A}]\rho=\{\hn H{-}\mathcal{A}H,\mathcal{A}\rho\}$ respectively [in the former case, (\ref{poisson}) shows that Poisson brackets satisfy a product rule]. Note furthermore that the commutator $[H,\mathcal{S}]$ simply does not occur in the classical theory.

As an afterthought to the above, we point at the possibility that superoperators could be declared points in a new space themselves, with hyperoperators mapping between them, and so on at infinitum. At least in the finite-dimensional case, the construction should proceed without difficulty. It is not clear to us to what extent this hierarchy has been pursued in mathematics.

\setlength{\unitlength}{1mm}
\begin{figure}[hb]
\begin{picture}(80,50)
  \put(10,50){\line(1,0){70}}
  \put(45,50){\line(0,-1){5}}
  \put(41,45){\line(1,0){8}}
  \put(42,44){\line(1,0){6}}
  \put(43,43){\line(1,0){4}}
  \put(10,16){\line(0,1){34}}
  \put(80,16){\line(0,1){34}}
  \put(7,30){\line(1,1){6}}
  \put(7,36){\line(1,-1){6}}
  \put(7,36){\line(1,0){6}}
  \put(10,16){\line(1,0){70}}
  \put(32,13){\line(1,1){6}}
  \put(38,13){\line(-1,1){6}}
  \put(65,13){\line(-1,1){6}}
  \put(59,13){\line(1,1){6}}
  \put(49,16){\line(0,-1){6}}
  \put(46,10){\line(1,0){6}}
  \put(46,9){\line(1,0){6}}
  \put(49,9){\line(0,-1){4}}
  \put(45,33){\circle*{1}}
  \put(45,33){\circle{2}}
  \put(52,45){$\phi\equiv0$}
  \put(14,32){$C_\mathrm{b}$}
  \put(30,20){$E_1,C_1$}
  \put(57,20){$E_2,C_2$}
  \put(53,8){$C_\mathrm{g}$}
  \put(47,1){$V_\mathrm{g}$}
  \put(13,18){$Q_\mathrm{b}$}
  \put(48,18){$Q_\mathrm{s}$}
  \put(47,32){$\phi_\mathrm{e}$}
\end{picture}
\caption{A loop consisting of a $d$--$d$ junction and a SET, threaded by an external flux.}
\label{fig}
\end{figure}
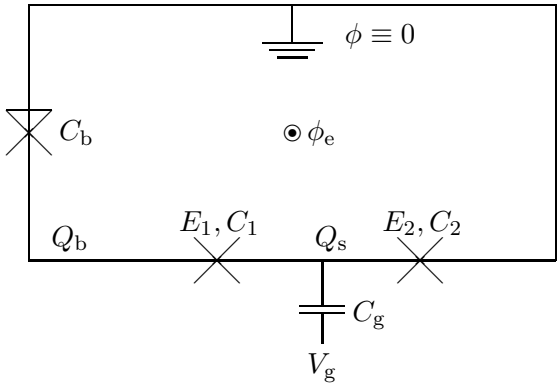

\section{Liouville-symmetric qubit circuit}
\label{circuit}

Let us give an example, where \L\nobreakdash-symmetry occurs naturally and actually facilitates practical calculations. Consider the circuit in Figure~\ref{fig}, containing an intrinsically degenerate phase qubit, coupled to a single-electron transistor (SET)~\cite{Tinkham} acting as a switch~\cite{PRSQ}. Readers not interested in Josephson devices can skip the circuit analysis and simply take its effective  Hamiltonian (\ref{Heff}) below as a postulated one, acting in a 4D Hilbert space. The exact nature of the qubit is immaterial here; it can for instance be realized as a $d$--$d$ grain boundary. A loop is formed by coupling the SET to the bulk of the qubit. The SET is taken superconducting as well, with Josephson couplings $E_{1,2}$, so that phase coherence can be maintained throughout the loop, and a phase frustration $\phi_\mathrm{e}$ can be imposed by an external magnetic flux. The SET and qubit phases are $\phi_\mathrm{s}$ and $\phi_\mathrm{b}$ respectively, with conjugate charges $Q_\mathrm{s}$ and $Q_\mathrm{b}$. The SET island can be polarized by a gate voltage $V_\mathrm{g}$, coupled through a gate capacitor $C_\mathrm{g}$. If the loop inductance is negligible and the SET (qubit\footnote{$C_\mathrm{b}$ also contains, and in fact may be dominated by, the qubit--SET lead capacitance.}) capacitances $C_{1,2}$ ($C_\mathrm{b}$) satisfy $C_1\ll C_\mathrm{b}$, standard electrostatics yields the Hamiltonian
\bea
  H&=&H_{\hn Q}+E_\mathrm{b}(\phi_\mathrm{b})-E_1\cos(\phi_\mathrm{b}{-}\phi_\mathrm{s})
    -E_2\cos(\phi_\mathrm{s}{-}\phi_\mathrm{e})\;,\label{Hfull}\\[2mm]
  H_{\hn Q}&=&\frac{1}{2}\begin{pmatrix} Q_\mathrm{b} &
    Q_\mathrm{s}{+}C_\mathrm{g}V_\mathrm{g}\end{pmatrix}
  \begin{pmatrix} C_\mathrm{b}^{-1} & C_1/(C_\mathrm{b}C_\Sigma) \\[1mm]
    C_1/(C_\mathrm{b}C_\Sigma) & C_\Sigma^{-1}\end{pmatrix}
  \begin{pmatrix} Q_\mathrm{b} \\[1mm] Q_\mathrm{s}{+}C_\mathrm{g}V_\mathrm{g}\end{pmatrix}\;,
  \nonumber
\eea
where $C_\Sigma\equiv C_1+C_2+C_\mathrm{g}$.

The qubit Josephson coupling $E_\mathrm{b}(\phi_\mathrm{b})$ is not specified in detail, except by stating that it has two degenerate minima at $\phi_\mathrm{b}=\pm\phi_0\ll2\pi$. If phase tunneling between them is small but not negligible, it is then appropriate to go to the qubit basis $\phi_\mathrm{b}=\phi_0\sB^z$, so that the free qubit Hamiltonian reduces to $H_\mathrm{b}\equiv Q_\mathrm{b}^2/(2C_\mathrm{b})+E_\mathrm{b}(\phi_\mathrm{b})\mapsto-\Delta\sB^x$, with $\Delta$ the tunneling amplitude. The remaining qubit operator occurring in (\ref{Hfull}) becomes $Q_\mathrm{b}\mapsto q\sB^y$, with $q\equiv i\bra{-\phi_0}Q_\mathrm{b}\ket{\phi_0}\in\mathbb{R}$. Note that $q$, being determined by tunneling, is of the order of $\Delta$, while e.g.\ $\sin\phi_0$ can be considerably larger. The derivation of a finite-dimensional effective Hamiltonian may be completed by making the two \emph{charge}-state approximation for the SET, i.e., $Q_\mathrm{s}\mapsto Q_0+e(1{+}\sS^z)$, upon which $e^{\pm i\phi_\mathrm{s}}\mapsto\sS^\pm\equiv(\sS^x\pm i\sS^y)/2$. Elementary manipulations lead to
\bea
  H_\mathrm{eff}&=&-\Delta\sB^x
    +\frac{C_1q\tilde{Q}}{C_\mathrm{b}C_\Sigma}\hp\sB^y
    -\frac{E_1\cos\phi_0{+}E_2\cos\phi_\mathrm{e}}{2}\hp\sS^x
    -\frac{E_2\sin\phi_\mathrm{e}}{2}\hp\sS^y
    +\frac{e\tilde{Q}}{C_\Sigma}\hp\sS^z\nonumber\\ &&{}
    +\frac{C_1qe}{C_\mathrm{b}C_\Sigma}\hp\sB^y\sS^z
    -\frac{E_1\sin\phi_0}{2}\hp\sB^z\sS^y\label{Hphys}\\[2mm]
  &\equiv&\alpha\sB^x+\beta\sB^y+\gamma\sS^x+\delta\sS^y
          +\epsilon\sS^z+\zeta\sB^y\sS^z+\eta\sB^z\sS^y\;,
\eeal{Heff}
where $\tilde{Q}=Q_0+e+C_\mathrm{g}V_\mathrm{g}$ is the effective charge offset from degeneracy.

\section{Hamiltonian analysis}

Focus on the case $\tilde{Q}=0$, where the SET is degenerate, i.e., acts as an open switch. By Eq.~(\ref{Heff}), this implies $\beta=\epsilon=0$.\footnote{In fact, the vanishing of $\epsilon$ is not essential, so that the analysis could be done for any $\tilde{Q}$ if the bit--SET \emph{charge} coupling $\propto C_1q$ is negligible.\label{note3}} Then, the quartic equation for the eigenenergies $\omega$ reduces\footnote{This happens whenever $\beta\epsilon\zeta=0$.} to a quadratic equation for $\omega^2$, with solutions
\beq
  \om^2=\alpha^2+\gamma^2+\delta^2+\zeta^2+\eta^2
    \pm2\sqrt{(\alpha\gamma{-}\zeta\eta)^2+\delta^2(\alpha^2{+}\eta^2)}\;;
\eeql{ef}
labeling $\om_1<\om_2<\om_3<\om_4$, these have precisely the difference property~(\ref{om-diff}). The corresponding eigenvectors [in the basis $\ket{\sB^z\sS^z}=(\ket{++},\ket{+-},\ket{-+},\ket{--})$] are
\beq
  \ket{n}=\begin{pmatrix} u\om_n \\ u[\gamma+i(\eta{+}\delta)]+v_n(i\zeta{-}\alpha) \\
    u(i\zeta{-}\alpha)+v_n[\gamma+i(\eta{-}\delta)] \\ v_n\om_n \end{pmatrix}\;,
\eeql{ev}
where $u=2(\alpha\gamma-\zeta\eta-i\alpha\delta)$ and $v_n=\alpha^2+\gamma^2+(\delta{+}\eta)^2+\zeta^2-\om_n^2$. Their norms are $\langle n|n\rangle=4\om_n^2v_n(v_n-2\delta\eta)$.

The pairing of states $(\ket{n},\ket{\bar{n}})$, with $\om_{\bar{n}}=-\om_n$, is readily explained by $\{\hn H,S\}=0$ for $S=\sB^z\sS^z$. For the vectors (\ref{ev}), one has $S\ket{n}=-\ket{\bar{n}}$.

To study qubit evolution in the presence of the SET, one can define the correlator ($\sigma^0\equiv\openone$, $j\neq0$)
\beq
  f_{jk\ell}(t)\equiv\Tr\left\{\sB^j(t)\sB^k\sS^\ell\right\}\;.
\eeq
Introducing the parity of the Pauli operators
\beq
  S\sigma_\mathrm{b/s}^j=p^j\sigma_\mathrm{b/s}^jS\quad\Rightarrow\quad p^0=-p^x=-p^y=p^z=1\;,
\eeq
one has
\bea
  f_{jk\ell}(t)&=&\Tr\left\{S^2e^{iHt}\sB^je^{-iHt}\sB^k\sS^\ell\right\}\nonumber\\
  &=&p^jp^kp^\ell\Tr\left\{Se^{-iHt}\sB^je^{iHt}\sB^k\sS^\ell S\right\}\nonumber\\
  &=&p^jp^kp^\ell f_{jk\ell}(-t)\;,
\eea
implying
\begin{widetext}
\beq
  f_{jk\ell}(t)=\left\{\begin{array}{ll} \ds\sum_{nm}\frac{\re[(\sB^j)_{nm}(\sB^k\sS^\ell)_{mn}]}
    {\langle n|n\rangle\langle m|m\rangle}\cos(\om_{mn}t)\;, & p^jp^kp^\ell=1\;; \\[5mm]
  \ds\sum_{n\neq m}\frac{\im[(\sB^j)_{nm}(\sB^k\sS^\ell)_{mn}]}
    {\langle n|n\rangle\langle m|m\rangle}\sin(\om_{mn}t)\;, & p^jp^kp^\ell=-1\;.\end{array}\right.
\eeql{fcases}
\end{widetext}
Here, $(a)_{nm}=\bra{n}a\ket{m}$ and $\om_{mn}=\om_m-\om_n$. Easy properties are $f_{jk\ell}(0)=4\delta_{jk}\delta_{\ell0}$, $|f|\le4$ in general, and $\dot{f}_{jk\ell}(0)=0$ if $[\sB^j,\sB^k]=0$. To dispose of trivialities upfront, note that since the identity matrix does not evolve, $f_{j00}(t)=\Tr\sB^j=0$.

Of the remaining 45 cases, an exhaustive search, using computer algebra to evaluate (\ref{fcases}) in terms of the states (\ref{ef})--(\ref{ev}), shows that no less than 25 cancel:
\bea
  f_{yx0}&=&f_{yy\ell}=f_{yz\ell}=f_{y0\ell}=f_{zx0}=f_{zy\ell}\nonumber\\
  &=&f_{zz\ell}=f_{z0\ell}=f_{xy0}=f_{xz0}=f_{xx\ell}=0\qquad(\ell\neq0)\;.
\eeal{cancel}
While (\ref{fcases}) can be simplified further using, e.g., ${(\sB^z)}_{n\bar{n}}={(\sS^z)}_{nn}={(\sB^y)}_{nn} ={(\sB^y)}_{n\bar{n}}={(\sB^x\sS^z)}_{nn}={(\sB^x\sS^z)}_{n\bar{n}}=0$ [by the \emph{conventional} symmetry $A=\delta\eta\sB^z+(\alpha\gamma{+}\zeta\eta)\sB^x\sS^x+\alpha\delta\sB^x\sS^y$---a deformed SET charge conjugation], these do not account for (\ref{cancel}) in a systematic, unified way. Apparently, the above expansion in eigenstates, while yielding explicit expressions for the nonzero correlators, is not suited to explain the vanishing of the other ones.

\section{Liouvillian analysis}
\label{liouville}

The above is to be contrasted with the Liouville approach. Let us return to the full Hamiltonian~(\ref{Heff}), and in line with its representation $H=\sum_{jk}H_{jk}\,\sB^j\otimes\sS^k$ also write $\rho=\sum_{jk}\rho_{jk}\,\sB^j\otimes\sS^k$. This generalizes the widely used Bloch vector for two-level systems~\cite{meystre}. Of course, $\rho_{00}=\frac{1}{4}\Tr\rho$ is conserved; for the other elements, the standard commutation relations for the Pauli matrices yield (\ref{neumann}) as
\begin{widetext}
\vspace{\abovedisplayskip}
\setlength{\unitlength}{.1mm}
\noindent\begin{picture}(1640,790)
\put(0,410){\parbox{\textwidth}{%
\beq
  d_t\!\begin{pmatrix} \rho_{y0}\\ \rho_{z0}\\ \rho_{xx}\\ \rho_{xy}\\ \rho_{xz}\\ \rho_{x0}\\
    \rho_{yx} \\ \rho_{yy} \\ \rho_{yz} \\ \rho_{zx} \\ \rho_{zy} \\ \rho_{zz} \\ \rho_{0x} \\
    \rho_{0y} \\ \rho_{0z} \end{pmatrix}=2\!
  \begin{pmatrix}
  \m0& -\alpha& \m0& \m\eta& \m0& \m0& \m0& \m0& \m0& \m0& \m0& \m0& \m0& \m0& \m0\\
  \m\alpha& \m0& \m0& \m0& -\zeta& -\beta& \m0& \m0& \m0& \m0& \m0& \m0& \m0& \m0& \m0 \\
  \m0& \m0& \m0& -\epsilon& \m\delta& \m0& \m0& \m0& \m0& \m\beta& \m0& \m0& \m0& \m0& \m0 \\
  -\eta& \m0& \m\epsilon& \m0& -\gamma& \m0& \m0& \m0& \m0& \m0& \m\beta& \m0& \m0& \m0& \m0 \\
  \m0& \m\zeta& -\delta& \m\gamma& \m0& \m0& \m0& \m0& \m0& \m0& \m0& \m\beta& \m0& \m0& \m0 \\
  \m0& \m\beta& \m0& \m0& \m0& \m0& \m0& -\eta& \m0& \m0& \m0& \m\zeta& \m0& \m0& \m0 \\
  \m0& \m0& \m0& \m0& \m0& \m0& \m0& -\epsilon& \m\delta& -\alpha& \m0& \m0& \m0& -\zeta& \m0 \\
  \m0& \m0& \m0& \m0& \m0& \m\eta& \m\epsilon& \m0& -\gamma& \m0& -\alpha& \m0 & \m\zeta& \m0& \m0\\
  \m0& \m0& \m0& \m0& \m0& \m0& -\delta& \m\gamma& \m0& \m0& \m0& -\alpha& \m0& \m0& \m0 \\
  \m0& \m0& -\beta& \m0& \m0& \m0& \m\alpha& \m0& \m0& \m0& -\epsilon& \m\delta& \m0& \m0& \m\eta\\
  \m0& \m0& \m0& -\beta& \m0& \m0& \m0& \m\alpha& \m0& \m\epsilon& \m0& -\gamma& \m0& \m0& \m0 \\
  \m0& \m0& \m0& \m0& -\beta& -\zeta& \m0& \m0& \m\alpha& -\delta& \m\gamma& \m0& -\eta& \m0& \m0\\
  \m0& \m0& \m0& \m0& \m0& \m0& \m0& -\zeta& \m0& \m0& \m0& \m\eta& \m0& -\epsilon& \m\delta \\
  \m0& \m0& \m0& \m0& \m0& \m0& \m\zeta& \m0& \m0& \m0& \m0& \m0& \m\epsilon& \m0& -\gamma \\
  \m0& \m0& \m0& \m0& \m0& \m0& \m0& \m0& \m0& -\eta& \m0& \m0& -\delta& \m\gamma & \m0
  \end{pmatrix}\!
  \begin{pmatrix} \rho_{y0} \\ \rho_{z0} \\ \rho_{xx} \\ \rho_{xy} \\ \rho_{xz} \\ \rho_{x0} \\
    \rho_{yx} \\ \rho_{yy} \\ \rho_{yz} \\ \rho_{zx} \\ \rho_{zy} \\ \rho_{zz} \\ \rho_{0x} \\
    \rho_{0y} \\ \rho_{0z} \end{pmatrix}.
\eeql{Lmat}}}
  \multiput(300,544)(59,0){20}{\line(1,0){40}}
  \multiput(670,56)(0,58){13}{\line(0,1){39}}
\end{picture}

\vspace{\belowdisplayskip}
\end{widetext}
The antihermiticity of the operator \L\ implicit in (\ref{Lmat}) is an easy consequence of the commutation with a Hermitian $H$ in (\ref{neumann}), see (\ref{anti-h}). For $\beta\rightarrow0$, one sees that \L\ decomposes into an upper $5\times5$ and a lower $10\times10$ block. This immediately explains the cancellations of $f$ [and many others, if one e.g.\ also studies $\Tr\{\sB^j(t)\sS^m(t)\sB^k\sS^\ell\}$] noted below (\ref{fcases}). In NMR, a Liouvillian analysis is sometimes also known as the ``direct method,'' especially when determining the spectrum of \L; the correspondence of the latter's eigenvalues to excitation frequencies bears some resemblance to Green's function methods~\cite{B&P}. While block-diagonal Liouvillians have been observed previously~\cite{beck}, these have seldom been pursued as due to a \emph{bona fide} physical symmetry, which is not necessarily linked to damping or decoherence.

Let us momentarily return to the detailed Hamiltonian~(\ref{Hphys}). For $E_1=0$, it describes a flux qubit coupled to a fluctuating charge. An advantage of flux qubits is that they can be designed insensitive to charge noise. Here, weak decoherence is due to the ``transition dipole moment''~$q$. For an interpretation, note that the qubit capacitor will have a voltage across it during flux jumps, which may well couple to neighboring charges; cf.~\cite{tian}. Presently, the usual charge-noise insensitivity argument only predicts that the qubit stays coherent if $E_1=q=0$, a correct but trivial statement for $H$ as in (\ref{Hphys}) with only $\alpha,\gamma,\delta,\epsilon\neq0$ (see Section~\ref{further} for more on uncoupled subsystems). We stress that our discussion, leading to~(\ref{Lmat}), concerns a different phenomenon; cf.\ note~\ref{note3}.

One experimental consequence may be formulated as follows. Start with the qubit pseudospin polarized in the $y$--$z$ plane, $\rho(0)=\half(\openone+\sB^yr\cos\theta+\sB^zr\sin\theta)\otimes\rho_\mathrm{s}$ ($0\le r\le1$). If at any later time $t$ one measures this spin along any direction in the same plane, the outcome distribution is independent of the SET's initial state $\rho_\mathrm{s}$, even though the two systems may interact strongly and anisotropically. This statement, rigorously proven above, does not seem to follow from any conventional consideration of this simple two-spin system.

In detail, the \L-symmetry uncovered above reads
\beq
  [\L,\mathcal{P}]=0
\eeql{LP}
[cf. (\ref{U-S})], where $\mathcal{P}=(\mathcal{P}_\mathrm{b}^x-\mathcal{P}_\mathrm{s}^0)^2$ is the projector onto the subspace spanned by the first five components of $\rho$ (plus the trivial normalization $\rho_{00}$) in (\ref{Lmat}). Explicitly:\footnote{Contrast with the projector describing a measurement of $\sB^x$, viz., $\mathcal{M}_\mathrm{b}^x(\rho)=\half(\rho+\sB^x\rho\sB^x)$, which retains both the $\sB^x$- \emph{and} the $\openone_\mathrm{b}$-components.} $\mathcal{P}_\mathrm{b}^x(\rho)=\frac{1}{2}\sB^x\otimes\Tr_\mathrm{b}\{\sB^x\rho\}$ etc. Inside each block, \L\ can of course be diagonalized separately; this would \emph{not} have been possible using an eigenbasis of the form (\ref{L-ev}), (\ref{ev}).

\section{Further examples and discussion}
\label{further}

In Sections \ref{circuit}--\ref{liouville}, the discussion has focused on density matrices and the von Neumann equation. However, (\ref{LP}) shows that what really matters is having an operator (there a projector~$\mathcal{P}$) commuting with the Liouvillian, as is also apparent in the general treatment of Section~\ref{theory}. Therefore, the whole discussion can be equivalently formulated in terms of the Heisenberg equation, $\dot{A}=-\L(A)$ for any observable $A$. This shows that the relevance of \L-symmetry is not in any way limited to quantum statistics.

Likewise, attention has been implicitly restricted to time-independent Hamiltonians. Now suppose that the parameters in (\ref{Heff}) have an arbitrary time dependence, except for $\beta(t)\equiv0$. The derivation of (\ref{Lmat}) remains valid, so that all cancellations~(\ref{cancel}) still hold. However, the simple formulas (\ref{fcases}) no longer apply, so that in the Schr\"odinger approach, even brute-force analytical verification of these cancellations is hampered.

Thinking of this situation as interaction with a \emph{classical} external field, one is naturally led to the next generalization. Namely, if the system is coupled to a quantum-mechanical ``outside,'' then in the Markov approximation this is reflected by the presence of nonunitary terms in \L, so that a Hamiltonian counterpart does not exist from the outset. The important point is that such a dissipative Liouvillian $\L_\mathrm{d}$ may still preserve the block structure of~(\ref{Lmat}), in which case the \L-symmetry will survive even in the presence of damping. Examples can be immediately given by replacing some of the parameters (other than $\beta$) in (\ref{Heff}) with e.g.\ oscillator-bath operators, so that the corresponding qubit--SET operators become error generators. It is not relevant for the argument whether the reduced evolution in qubit--SET Liouville space (\ref{Lmat}) is Markovian or not. Physically, we think of a situation in which the circuit Hamiltonian \emph{retains} its form (\ref{Hfull}) (at least to leading order in some damping parameter), but with e.g.\ $\phi_\mathrm{e}$ susceptible to (quantum) noise. Intuitively, quantum information stored in one of the blocks does not ``leak'' to other ones, comparable to the behavior of quantum error-correcting codes~\cite{dec-subs}. Pending a comprehensive theory of \L-symmetric open systems, the above re-opens the question whether the notion of decoherence-free subsystems indeed is ``ultimate''~\cite{dec-exp}.

Of particular interest would be the case when $\L_\mathrm{d}$ is nonzero in only \emph{one} of the blocks, so that the density-matrix components in the other block do not feel its influence at all. This latter block could then rightly be called a \emph{decoherence-free Liouville subspace}, in which quantum information can be stored error-free, provided that the unitary (internal) part of \L\ is under sufficient control. In fact, it can be precisely indicated where our discussion deviates from, e.g., the---correct---analysis in~\cite{dec-free}. The latter's theorems 1 and~2 state, in the present terminology, that to every decoherence-free Hilbert subspace $C_N$ one can associate a decoherence-free Liouville subspace $\mathcal{M}_N$; we ask if there are $\mathcal{M}$ not of this form. Note that the basis vectors of $\mathcal{M}$ may well be ``off-diagonal elements'' in the usual parlance. Therefore, we have insisted on $\mathcal{P\hn L}_\mathrm{d}=\L_\mathrm{d}\mathcal{P}=0$, where $\mathcal{P}$ denotes the projector on $\mathcal{M}$. In other words, it is not sufficient to have\footnote{However, the corresponding condition for the \emph{Hilbert}-space projection $P$ and error generators $F_\alpha$, when these exist for conventional $\mathcal{M}_N$, will automatically lead to our stricter criterion for $\L_\mathrm{d}$. Notice further that in Liouville space we impose conditions both on $\mathcal{P\hn L}_\mathrm{d}$ and on $\L_\mathrm{d}\mathcal{P}$; in contrast, the Hilbert-space generators are Hermitian, so that always $PF_\alpha=F_\alpha P$.} $\mathcal{PL}_\mathrm{d}=\L_\mathrm{d}\mathcal{P}\propto\mathcal{P}$; this is to be contrasted with the non-semisimple case in~\cite{lidar}. First of all, however, it needs careful investigation whether this is possible in principle for time evolution which ultimately is Hamiltonian in the ``universe'' of system plus environment. That is, can there exist error generators which vanish in some unconventional Liouville subspaces $\mathcal{M}$ but not in others; cf.\ (\ref{Lmat}), where all parameters other than $\beta$ occur in \emph{both} blocks. In summary, one may anticipate interesting future research into \L-symmetric dissipative models.

The harmonic oscillator (HO) figures in almost any physics text, and the present is no exception. It is described by $H=\half(p^2+q^2)=a^{\dagger}a+\half$. In the notation of Section~\ref{theory}, set $\mathcal{S}_+\equiv a^{\dagger}_\mathrm{l}a_\mathrm{r\vphantom{l}}^\phd$ so that $\mathcal{S}_-\equiv a_\mathrm{l}^\phd a^{\dagger}_\mathrm{r\vphantom{l}}=\mathcal{S}_+^{\dagger}$. Then $[\L,\mathcal{S}_+]=[\L,\mathcal{S}_-]=0$, while e.g.\ $[H_\mathrm{l},\mathcal{S}_+]=\mathcal{S}_+\neq0$; all of these are conveniently evaluated in superoperator notation, using only the representation~(\ref{HlHr}) of \L\ and the calculation rules, together with the standard $[H,a]=-a$. $\mathcal{S}_\pm$ can also be studied in the context of the coherent~\cite{louisell,meystre}\linebreak and other interesting non-diagonal states~\cite{yuen} which the HO is known to possess. For instance, $\mathcal{S}_-(\ket{\alpha}\bra{\alpha})=|\alpha|^2\ket{\alpha}\bra{\alpha}$ for the coherent state $\ket{\alpha}\equiv\exp\{\alpha a^{\dagger}{-}\alpha^*a\}\ket{0}$. One also verifies $[\mathcal{S}_+,\mathcal{S}_-]=i\L$, so that the Lie algebra of $(\mathcal{S}_+,\mathcal{S}_-,\L)$ is isomorphic to the one of $(q,p,\openone)$. Note further that $a,a^{\dagger}$ are said to be SUSY operators in supersymmetric quantum mechanics~\cite{SUSY}. In fact, the HO with these operators is the prototypical system. We thus arrive at a connection between our superoperator symmetry and the conventional ``super'': the HO is its own superpartner because deleting the ground state yields the same spectrum, apart from a shift by $\om{=}1$---all its energy differences are degenerate.\label{super}

This example is of special interest because the classical limit can be studied in detail. The Liouvillian is $\L=q\pt_p-p\pt_q$, and the system's integrals are of the form $f\circ H$ for any real~$f$. The Hamiltonian symmetries are $e^{\phi(H)\L}$ for any real~$\phi$, including parity $e^{\pi\L}$. Observing that above $\mathcal{S}_+$ added one quantum of energy to the system, we set $\pt_{\!_H}\equiv p\pt_p+q\pt_q$, upon which $[\L,\pt_{\!_H}]=0$ is readily verified. However, obviously $[H,\pt_{\!_H}]\neq0$. This \L-symmetry is nontrivial: it expresses that nearby trajectories, \emph{not} on the same constant-energy contour, keep moving in phase. The transformations generated by $\pt_{\!_H}$ are not canonical, since they do not preserve phase-space volume. Explicitly, $A:(p,q)\mapsto\boldsymbol{(}\alpha(H)p,\alpha(H)q\boldsymbol{)}$ is associated with an \L-symmetry for monotonic $\alpha$.\footnote{One can consider wider possibilities by not requiring $A$ to be invertible. Allowing a singularity at the origin of phase space, exotics such as $\alpha(H)=H^{-1}$ are also possible; all of these lead to $[\L,\mathcal{A}]=0$.} This sheds considerable light on the \L-symmetry phenomenon: in quantum mechanics, only energy differences correspond to excitations and quantum beats, hence are observable. In classical mechanics, only the Hamiltonian \emph{flow} is observable, and \L-symmetries preserve it.

\begin{figure}[t]
\begin{picture}(74,82)
  \put(0,0){\includegraphics[scale=0.7]{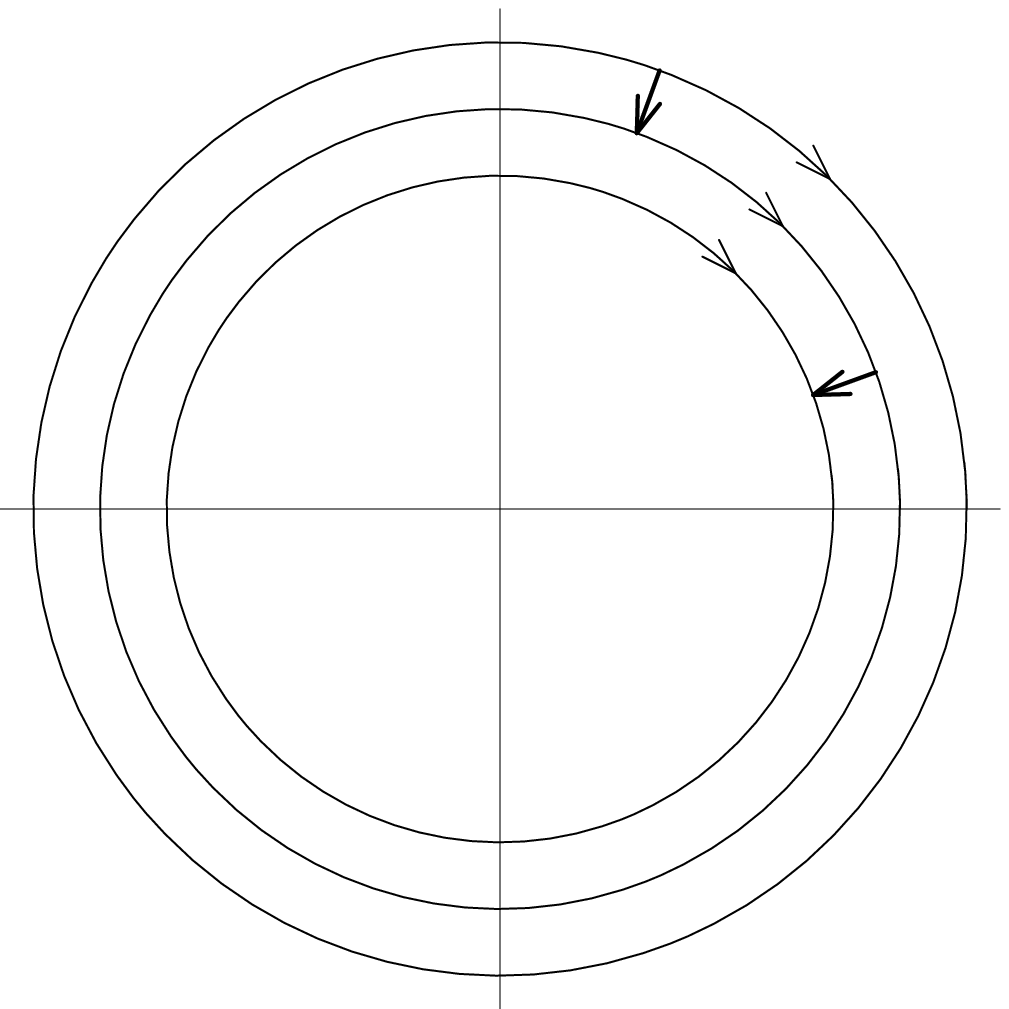}}
  \put(72,34){$q$}
  \put(34,74){$p$}
  \put(3,67){(a)}
\end{picture}
\qquad
\begin{picture}(74,82)
  \put(0,0){\includegraphics[scale=0.7]{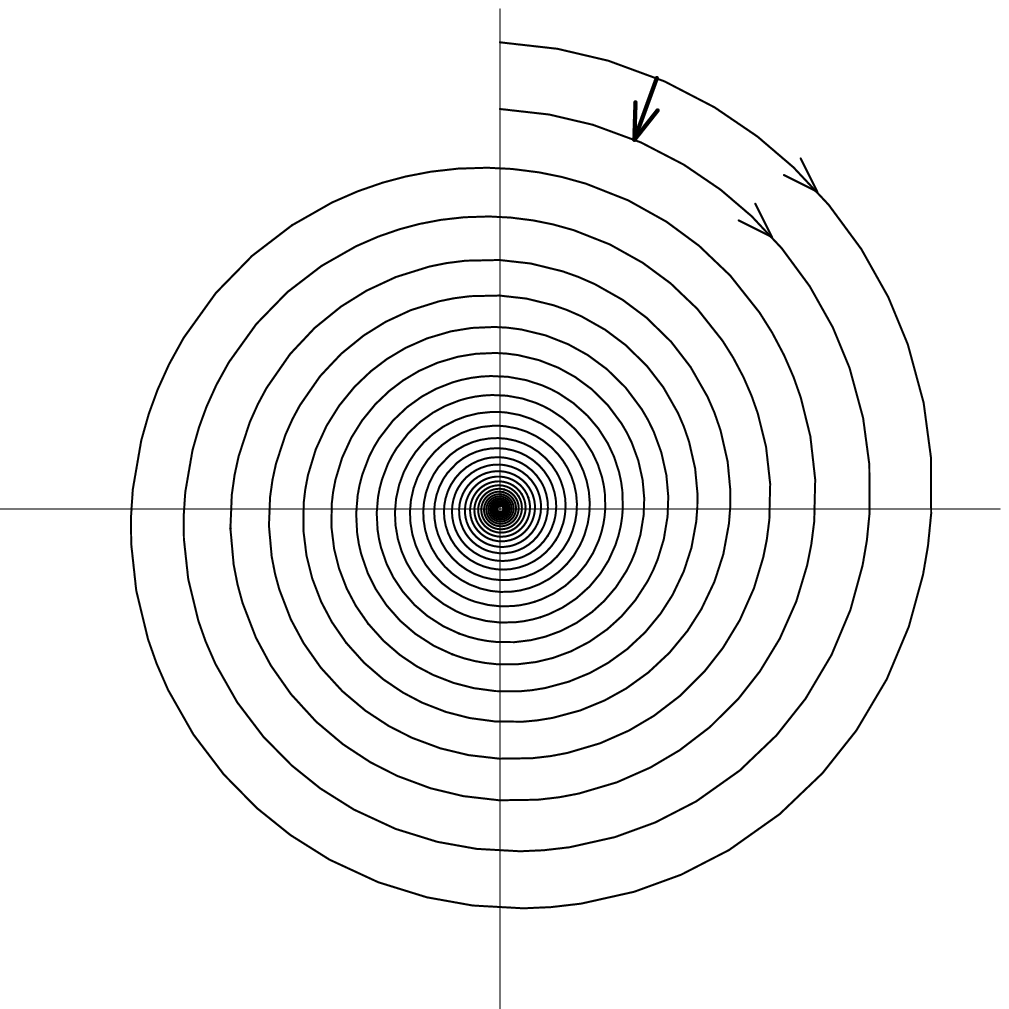}}
  \put(72,34){$q$}
  \put(34,74){$p$}
  \put(3,67){(b)}
\end{picture}
\caption{Phase-space flow for the damped-oscillator Liouvillian $\L_\mathrm{tot}=q\pt_p-p\pt_q+\gamma\pt_{\!_H}$, with $\gamma=0$~(a) and $\gamma=\frac{1}{20}$ (b) respectively. Bold arrows indicate the action of the \L-symmetry $\pt_{\!_H}$, which maps integral curves onto each other in \emph{both} cases.}
\label{fig2}
\end{figure}

Thus, \L-symmetry also plays a role in classical mechanics. Given that in quantum mechanics we contemplated the symmetry's survival under certain types of dissipation, one should look for a classical counterpart to that as well. In fact, for harmonic problems such a counterpart is well known to be \emph{additive} noise and \emph{velocity-independent} friction, which preserve the linearity of the equations of motion. For instance, $[\L_\mathrm{tot},\pt_{\!_H}]=0$ for all $\L_\mathrm{tot}=\L+\gamma p\pt_p+\mu q\pt_q$,\footnote{However, of the finite transformations $A$ considered in the previous paragraph, only those with $\alpha(H)=\text{const}$ are permissible if $\gamma$ and/or $\mu$ are nonzero.} which includes both physical Ohmic friction ($\mu=0$) and the \textit{ad hoc} $\L_\mathrm{tot}=\L+\gamma\pt_{\!_H}$ for $\mu=\gamma$, in which case the phase-space trajectories become simple logarithmic spirals; see Figure~\ref{fig2}. The eigenfrequencies read $\om=\pm\sqrt{1-(\gamma{-}\mu)^2\!/4}-\frac{i}{2}(\gamma{+}\mu)$. As a result of scaling symmetry, i.e., linearity, the problem remains tractable considerably beyond the case of a single oscillator~\cite{opa}, and can be given a description strikingly similar to the conservative case~\cite{rmp}. In view of the above, one arrives at the prediction that this situation has a parallel for certain \emph{non}harmonic systems, provided that the damped/diffusive Liouvillians involved are compatible with the \L-symmetries of the conservative dynamics.

A variation on the HO is the Stark ladder, which has its Hilbert space spanned by $\{\ket{n}:n\in\mathbb{Z}\}$. In terms of the shift $a=\sum_n\,\ket{n}\bra{n{+}1}$ satisfying $aa^{\dagger}=a^{\dagger}a=\openone$, the Hamiltonian can be given as $H=\sum_nn\ket{n}\bra{n}-\Delta(a+a^{\dagger})$. Again $[H,a]=-a$, so that the definition of $\mathcal{S}_\pm$ and their commutators with $\L,H_\mathrm{l/r}$ can be immediately transcribed from the HO; however, presently $[\mathcal{S}_+,\mathcal{S}_-]=0$. The Stark ladder is prototypical of many problems with a source, including solids in an electric field and current- (voltage-)biased Josephson junctions in the phase (charge) representation~\cite{Tinkham}. In particular, if the ``shifting'' degree of freedom is shift-invariantly coupled to other ones, a nontrivial problem ensues. While $\mathcal{S}_\pm$ then still are exact \L-symmetries, these are difficult to exploit using conventional methods~\cite{MSG}.

The simplest possible case concerns two uncoupled subsystems. Then, one has $\mathbb{H}=\mathbb{H}^{(1)}\otimes\mathbb{H}^{(2)}$ for the Hilbert space and $H=H^{(1)}\otimes\openone^{(2)}+\openone^{(1)}\otimes H^{(2)}$ for the Hamiltonian. The spectrum can now be labeled as $\om_{ik}=\om_i^{(1)}+\om_k^{(2)}\Rightarrow\forall i,j,k,\ell:\om_{ik}-\om_{i\ell}=\om_{jk}-\om_{j\ell}$ [cf.~(\ref{om-diff})]. It is a staple of physics that finding such a decomposition of $(\mathbb{H},H)$ constitutes a drastic simplification of the problem. However, apparently it only leads to a proper symmetry on the Liouville level, as it is energy differences not energies which are degenerate. If $\mathcal{S}^{(1,2)}$ are \L-symmetries in the subsystems, then\footnote{As always, $\mathcal{S}$ is defined to act on entangled density matrices by superposition.} $\mathcal{S}(\rho^{(1)}\otimes\rho^{(2)})\equiv
\mathcal{S}^{(1)}(\rho^{(1)})\otimes\mathcal{S}^{(2)}(\rho^{(2)})$ is one in the composite system. The case of most interest is $\mathcal{S}^{(1)}=\openone$ and the ``forgetful'' $\mathcal{S}^{(2)}=\Tr$. The latter can be calculated without any knowledge about the second subsystem, and erases all reference to it. While for a single system this would merely be (proportional to) the projection on $\rho=\openone$, presently it leaves a meaningful problem for the first subsystem. We shall not elaborate on the classical counterpart, which is broadly similar.

\section{Conclusion}

We have introduced the concept of Liouville symmetry, of which we know no previous systematic and general treatment in the literature. Examples are neither excessively rare or contrived, nor is the symmetry's actual use unwieldy. Namely, one still deals with linear evolution and eigenvalue equations, albeit in a space larger than Hilbert space; for classical mechanics, it is standard to visualize the Liouville flow in phase space. In particular, there can be advantages to a Liouville formulation even for closed, unitary systems. We have only touched upon the application to open systems and quantum-information processing. For instance, the issues of storing, manipulating, and retrieving such information in/from the unconventional subspaces concerned, and the extent to which this is at all possible, are important and cannot be ignored. Also, group-theoretic aspects await elaboration. We hope that the present work may provide both the framework and a stimulus for such more detailed investigations.

In any case, all information stored in the matrix elements of $\rho$ is physical, and can be retrieved through, e.g., quantum-process tomography (QPT)~\cite{tomo}. For the Hamiltonian (\ref{Heff}), this should be eminently feasible using present-day technology, though perhaps sooner for nuclear spins than for high-$T_\mathrm{c}$ Josephson devices. With some urgency, we therefore propose an experiment which does exactly that: QPT for an \L-symmetric system.

\begin{acknowledgments}
We thank M.H.S. Amin, G.E.W. Bauer, A.~Blais, J.P. Hilton, D.A. Lidar, G.~Rose, M.F.H. Steininger, and L.G. Suttorp for discussions. The referees are thanked for their remarks, especially for pointing out Ref.~\cite{B&P}. AMB thanks the Chinese University of Hong Kong for its hospitality during the preliminary stage of this work.
\end{acknowledgments}

\end{document}